\documentclass[11pt]{iopart}
\usepackage{graphicx}

\begin{document}

\title{Dissipation in a quantum wire: fact and fantasy}
\footnote{See {\em AIP Conference Proceedings}, vol. 1063, pp 26-34 (2008).}

\author{Mukunda P. Das$^1$ and Frederick Green$^2$}
\address{$^1$ Department of Theoretical Physics, IAS,
The Australian National University, Canberra, ACT 0200, Australia.}
\address{$^2$ School of Physics, The University of New South Wales,
Sydney, NSW 2052, Australia.}

\begin{abstract}
Where, and how, does energy dissipation of electrical energy
take place in a ballistic wire?
Fully two decades after the advent of the transmissive
phenomenology of electrical conductance,
this deceptively simple query remains unanswered.
We revisit the quantum kinetic basis of dissipation
and show its power to give a definitive answer to our query.
Dissipation leaves a clear, quantitative trace in the
non-equilibrium current noise of a quantum point contact;
this signature has already been observed in the laboratory.
We then highlight the current state of accepted understandings
in the light of well-known yet seemingly contradictory measurements.
The physics of mesoscopic transport rests not in coherent carrier
transmission through a perfect and dissipationless
metallic channel, but explicitly in their dissipative inelastic
scattering at the wire's interfaces and adjacent macroscopic leads.
\end{abstract}


\section{State of Play}

In recent years the understanding of mesoscopic transport
has been rewritten through the insights of Landauer, B\"uttiker,
Imry, and others
\cite{p1,p3}.
Succinct and successful, this phenomenology of electron-wave
transmission characterizes mesoscopic current flow in terms of two effects:
the mismatch of carrier density between large metallic reservoirs
(the terminals) across which a low-dimensional conductor
(the quantum point contact) of actual interest is connected; and,
induced by the density mismatch,
lossless quantum transmission of single carriers through
the conductor (visualized as a potential barrier).

We are confronted with extremely small structures, possibly of molecular
size. Thus they experience a high degree of {\em openness} to their
macroscopic environment. A striking signature of transport
in such a quantum point contact is the
discretization of its conductance into ``Landauer steps'' in units of
$2e^2/h\approx0.078{\rm mS}$.

Landauer's conductance steps are explained via collisionless
single-electron quantum transmission through a one-dimensional
barrier. The process is evidently elastic; that is, loss-free.
However, simple quantum-coherent carrier transmission cannot
engage with the central issue of conduction: {\em What causes dissipation
in a ballistic quantum point contact}?

The question is far more than academic, and finding the
right answer to it is far more than an esoteric quest.
For, in the very near future -- if not right now -- reliable
and effective nano-electronic design will demand theoretical
treatments that are not, at their best, merely plausible
or cosmetic but credible physically and applicable practically.

It is beyond the scope of coherence-based phenomenologies to cover
the necessary physics in a substantial sense. The reason is simple:
in their very constitution, coherence methods are incapable
of characterizing inelasticity and energy dissipation.
So it is not surprising to see, in several classic
texts that expound coherent-transmission theory
\cite{p1,p3,kum},
little serious attempt to understand, and
come to grips with, the explicit microscopic action of dissipation.

Without a cogent description of dissipation,
{\em resistive} current flow makes no sense.
From the beginning, rather than from afterthought,
an answer has to be sought in the
manifest physical role of inelastic scattering,
and ballistic devices are no exception.
Such a description will connect with first principles to
provide a natural explanation for mesoscopic conductance
quantization, and much more besides.


In the first part of this paper we outline the answer to our question
within the established forms of many-body quantum kinetics
\cite{DG1}.
The microscopically based use of many-body methods leads not only
to ideal conductance quantization
in full account of inelastic energy loss
\cite{DG2},
but also resolves a long-standing experimental enigma
\cite{rez}
in the noise spectrum of a quantum point contact (QPC)
\cite{DG3}.
Because they follow naturally from microscopics,
the same developments foreshadow a systematic pathway to
the truly predictive design of novel and useful structures.

In the second part we discuss the idea of the ``intrinsic'' resistance
proper to a quantum wire, and how this important issue is understood
from the standpoint of quantum kinetic theory.
This topic is all the more necessary to address
because there exist well canvassed
experimental results that appear to defy the basic microscopic
understanding of the intrinsic mesoscopic resistance.

\section{The Physical Issue}

The core issue in the physics of conduction is plain to set out. Any
finite conductance $G$ must dissipate electrical energy at the rate
$P = IV = GV^2$, where $I = GV$ is the current and $V$ the potential
difference across the terminals of the driven conductor.
It follows that there must be an explicit physical mechanism
(emission of optical phonons is one example) by which the net
energy gained by carriers, when transported from source to drain,
is transferred irreversibly to the surroundings.
Alongside any elastic and coherent scattering processes,
inelastic processes must always be in place. When these are
harnessed together $G$ is uniquely determined; yet it is only
the energy-dissipating mechanisms that secure the
{\em thermodynamic stability} of steady-state conduction.

Since the early work of Callen and Welton, Kubo
and later P. C. Martin in the 'fifties, there has
been a complete microscopic understanding
of the universal power-loss formula $P = GV^2$
(see for example Refs. \cite{kubo} and \cite{wims}).
It resides in the fluctuation-dissipation theorem,
valid for {\em all} resistive devices at all scales,
in all circumstances. The theorem expresses
the requirement for thermodynamic stability.
With it comes the conclusion that
\cite{DG1,DG2}

\begin{itemize}
\item
inelasticity is necessary and sufficient to stabilize current
flow at finite conductance;

\item
ballistic quantum point contacts have finite $G \propto 2e^2/h$;
therefore

\item
the physics of energy loss is indispensable
to a proper theory of ballistic transport.
\end{itemize}

The physics of explicit inelastic scattering is beyond the scope 
of transport models that rely only on coherent quantum scattering
to explain the origin of $G$ in quantum point contacts. Coherence
implies elasticity, and elastic scattering is always loss-free:
it conserves the energy of the scattered particle. This reveals
the deficiency of purely elastic models of transmission. We now
review a well-defined microscopic remedy for this deficiency.

\section{The Physical Solution}

To allow for the energy dissipation vital to any microscopic
description of ballistic transport, we recall that open-boundary
conditions imply the intimate coupling of the QPC channel to its interfaces
with the reservoirs. The interface regions must be treated
as an integral part of the device model. They are the very sites
for strong scattering effects: {\em dissipative} many-body events
as the current  enters and leaves the ballistic channel,
and {\em elastic} one-body events as the carriers interact
with background impurities, the potential barriers that confine and funnel
the current, and so on. 

The key idea in our treatment is to subsume the interfaces within
the total kinetic description of the ballistic channel.
At the same time, strict charge conservation in an {\em open}
device requires the direct supply and removal of current
by an external generator
\cite{wims}.
Thus the current cannot depend on the physics
of the local reservoirs. This canonical requirement sets the
quantum kinetic approach entirely apart from Landauer-like treatments
\cite{p1},
which rest upon the phenomenological notion that the
current depends on density differences between reservoirs.

\begin{figure}
%
\center{\includegraphics[height=80mm]{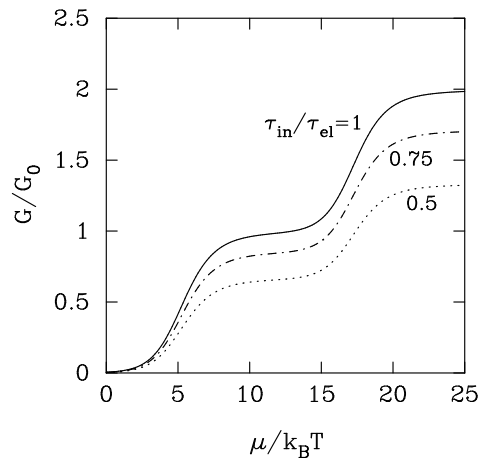}}
%
\vspace{-15mm}
\caption
{Conductance quantization in a two-band
ballistic point contact, as a function of chemical
potential $\mu$, calculated with our kinetic theory~\cite{DG1}.
Full curve: ideal ballistic channels.
Broken curves: non-ideal behaviour increases
with the onset of inelastic phonon emission inside the contact.}
\label{fig1}
\end{figure}

\subsection{Ballistic Conductance}

It is straightforward to write the algebra for the conductance
in our model system. (For a more detailed description see
Ref. \cite{DG1}.)
A uniform, one-dimensional ballistic QPC, of
operational length $L$, will be associated with two mean free paths
determined by $v_{\rm F}$, the Fermi velocity of the electrons,
and a pair of characteristic scattering times
$\tau_{\rm el},\tau_{\rm in}$. Thus

\begin{equation}
\lambda_{\rm el} = v_{\rm F}\tau_{\rm el};{~~}
\lambda_{\rm in} = v_{\rm F}\tau_{\rm in}.
\label{e1}
\end{equation}

\noindent
Respectively, these are the scattering lengths set by the
elastic and inelastic processes active at both interfaces.
From the viewpoint of measurable scattering behaviour,
each length encodes the same information as its time.

The device (the QPC {\em and} its interfaces) has a conductive
core that is collisionless. It follows that

\begin{equation}
\lambda_{\rm el} = L = \lambda_{\rm in}.
\label{e2}
\end{equation}

\noindent
Finally, the channel's conductance is given by the familiar formula

\begin{equation}
G = {ne^2\tau_{\rm tot}\over m^*L}
= {2k_{\rm F}\over \pi}{e^2\over m^*L}
{\left( {\tau_{\rm in}\tau_{\rm el} \over
         {\tau_{\rm el} + \tau_{\rm in}}} \right)};
\label{e3}
\end{equation}

\noindent
the effective mass of the carriers is $m^*$.
In the first factor of the rightmost expression for $G$
we rewrite the density $n$ in terms of the Fermi momentum $k_{\rm F}$;
in the final factor, we use Matthiessen's rule
$\tau_{\rm tot}^{-1} = \tau_{\rm el}^{-1} + \tau_{\rm in}^{-1}$
for the total scattering rate in the system.

On making use of equations (\ref{e1})--(\ref{e3}), the conductance
reduces to

\begin{equation}
G = 2{e^2\over \pi\hbar}{\hbar k_{\rm F}\over m^* L}
{\left( {(L/v_{\rm F})^2 \over 2L/v_{\rm F}} \right)}
= {2e^2\over h} \equiv G_0.
\label{e4}
\end{equation}

\noindent
This is exactly the Landauer conductance of a
single, one-dimensional, ideal channel.

In Figure 1 we plot the results of our model for a QPC
\cite{DG1}
made up of two one-dimensional conduction sub-bands with threshold
energies energies set at
5$k_{\rm B}T$ and 17$k_{\rm B}T$, in thermal Boltzmann
units $k_{\rm B}T$.
We have used the natural extension of equation (\ref{e4}) to the case
where more than one channel may open up to conduction, depending on
temperature $T$
and the size of the chemical potential $\mu$. As we increase
the role of inelastic scattering by making $\tau_{\rm in}$
shorter than $\tau_{\rm el}$ the conductance duly
undershoots its ideal, ballistic Landauer limit. But its step
structure survives.

None of the singular assumptions, commonly adduced
to explain conductance quantization as coherent transmission
\cite{p1,p3},
is needed by our conserving microscopic approach.
Indeed, the result emerges naturally from
completely standard quantum kinetics.

What is the crux of the logic of equation (\ref{e4})? It is
precisely the manifest and indispensable role of inelastic energy loss,
acting in concert with elastic scattering and on a {\em completely
equal physical footing}.
That its action is one of the underpinnings of quantum transport
is expressed in the fluctuation-dissipation theorem
\cite{kubo}.

Charge conservation -- the other and concurrent underpinning -- is
always guaranteed if one applies proper microscopically
consistent open-boundary conditions at the interfaces
\cite{DG1,DG2,wims}.
These twin canonical requirements, imposed by physics
and not by anything else, have so far not been evident
within heavily phenomenological derivations of equation (\ref{e4}).
It remains a challenge for the alternative phenomenologies
of ballistic conductance to demonstrate their direct
provenance from microscopics, at least to the same level of
logical consistency as the quantum kinetic analysis.

\subsection{Non-equilibrium Noise}

The noise response of a quantum point contact is a fascinating
aspect of mesoscopic transport, and a more demanding one both
experimentally and theoretically. It is in the noise characteristics
of a ballistic wire that the predictive capacity of a quantum
kinetic description comes really to the fore.

\begin{figure}
\center{\includegraphics[height=65mm]{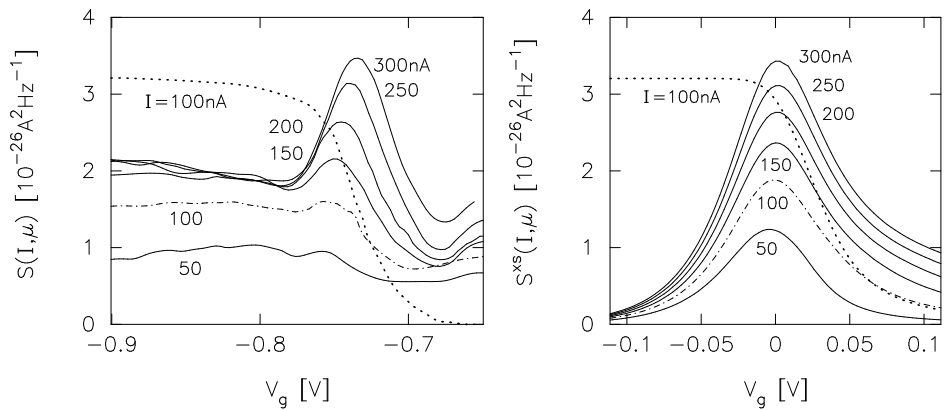}}
\caption
{Non-equilibrium current noise of a QPC at constant
source-drain current, as a function of gate bias. Left: data
from Reznikov {\em et al.}~\cite{rez}. Right: calculation
from Green {\em et al.}~\cite{DG3}. In each case the
dotted line traces the standard shot-noise prediction at 100nA
using, as respective inputs, measured and calculated data for $G$.
The standard prediction is well wide of the mark.}
\label{fig2}
\end{figure}

In 1995, a landmark measurement
of non-equilibrium noise was performed by the Weizmann group
\cite{rez},
which yielded a very puzzling result. Whereas conventional models
\cite{p3}
predicted a strictly monotonic shot-noise signal for a QPC driven
at constant current levels, the data showed a series of very strong
peaks at threshold (where the carrier density in the QPC rapidly
grows and becomes metallic). This is in marked contradiction to
theoretical expectations.

Remarkable as they still are, the Weizmann results remained
absolutely unexplained for a decade.
We have now accounted for them within our strictly
conserving kinetic description
\cite{DG3}.

In Fig. 2 we display the experimental data side by side with our
computation of excess QPC noise under the same conditions
\cite{DG3}.
There is close accord between the measured constant-current
peaks and our calculation. This
is in obvious contrast to standard phenomenology
\cite{p3}
which predicts no peaks whatsoever in this situation.

It is worth noting that the additional measurements of
noise-peak structures carried out at constant source-drain voltage
\cite{rez}
(a more common experimental practice)
are as well described by our quantum kinetics
as they are by any coherent-transmission model.
Furthermore, our calculation
\cite{DG3}
also shows a substantial monotonically rising noise background
at higher gate voltages.
This feature is quantitatively consistent with observations,
and again one that is completely absent from
phenomenological models of the same data
\cite{p3}.

\section{Intrinsic Resistance of a Quantum Wire: Fact or Fantasy?}

\subsection{Perfect Conductor}

Before analyzing the intrinsic resistance we stress once more
the central physical property of the Landauer formula deduced
from quantum kinetics: {\em it demands quantitative inclusion
of inelastic scattering if it is to emerge at all naturally}.
In the foregoing discussion of equation (\ref{e4}) we have treated
resistance as the total resistance of the mesoscopic device in
series with the access resistance of the interfaces and
macroscopic leads, the latter being the actual reference contacts
for external measurements.

The ``intrinsic'' resistance of a mesoscopic device should therefore
exclude all the access resistances: not only those belonging
to the external apparatus but, crucially, those of the 
boundaries with the leads. We do not have space here to cover
fully the quantum kinetic approach to intrinsic resistance.
It is the outcome of the system's overall response
to an external voltage, by which the field internal to the device
is minimized through self-consistent electrostatic screening
at the boundaries.
A detailed microscopic account may be found,
for example, in the study by Kamenev and Kohn
\cite{kk}
who apply standard first-principles methodology
\cite{kubo}
to the problem.

A practical way to characterize the intrinsic
(in some real sense, actual) device
resistance is to attach non-invasive voltage probes across the structure
in such a way that local current flow is undisturbed by any
measurement of the voltage drop between the probes.
The set-up of two current probes at the macroscopic leads,
augmented with two non-invasive voltage probes as close
as possible to the device, is referred to as a ``four-terminal'' arrangement.
Provided one can quantify the degree of disturbance of the current
by the inner terminals, this approach gives practical meaning to
theoretical discussion of the real resistance of a mesoscopic wire.

A striking example of a four-terminal measurement was given by
de Picciotto {\em et al}.
\cite{depic}
In this experiment, both two- and four-terminal resistances were
measured for an essentially ballistic, one-dimensional conducting
channel with inner probes designed to minimally disturb the
current in the device.

From two-terminal resistance measurements
(without internal probing; device and leads configured in series)
de Picciotto {\em et al}. obtained strong quantized steps in the
resistance, close to the ideal Landauer prediction for a ballistic
quantum channel; close but not quite coincident,
for the steps all show a clear shortfall of about 7\%
from ideal quantization. On the other hand, the intrinsic
resistance of the same structure was measured to be vanishingly small.

What is the interpretation of this experiment? First, in its
two-terminal version the Landauer formula tells us that, given
probability ${\cal T}$ for electron-wave transmission through
the channel, the source-drain resistance for a single occupied sub-band is

\begin{equation}
R_2 = {h\over 2e^2} {1\over {\cal T}}.
\label{r2} 
\end{equation}

\noindent
In an ideal case ${\cal T} = 1$. If the measured deviation in $R_2$ is a few
percent, one can infer a non-ideal ${\cal T} < 1$. Second, the four-terminal
resistance -- intrinsic to the device alone -- is
\cite{pwa}

\begin{equation}
R_4 = {h\over 2e^2} {\left( {1\over {\cal T}} - 1 \right)}.
\label{r4}
\end{equation}

\noindent
B\"uttiker
\cite{butt}
has derived a more general four-terminal relation in terms of the
partial transmission amplitudes between any one probe and any other.
When invasive influences are small, as in actual measurements,
this formula can be shown to be equivalent to equation (\ref{r4})
\cite{ferry}.

According to the four-terminal result of Ref. \cite{depic},
namely that the intrinsic resistance $R_4$ vanishes, it follows
that ${\cal T}$ is unity.
There is no intrinsic resistance here. {\em A fortiori} there is
none that can be said to be quantized.

In another experiment Reilly {\em et al.}
\cite{reilly}
also measured the four-terminal conductance ($1/R_4$)
of a ballistic conductor. Their raw data exhibit a perfect
Landauer staircase as a function of voltage applied
via a side gate to modulate the channel width. 
For this result Reilly {\em et al.} stress that
``no attempt has been made to adjust the plateau heights to
fit with quantized units of $2e^2/h$.'' That is: (i) $R_4$ is
free of artefacts and (ii) the intrinsic {\em conductance} is
decidedly finite and so is $R_4$.

Regarding the Landauer formulae (\ref{r2}) and (\ref{r4}),
one immediately faces a difficulty with respect
to their universal and mutually consistent understanding.
For, in the literature we are being offered two completely
different observations of $R_4$ (the nominally intrinsic
resistance of a clean quasi-one-dimensional quantum wire):
in one measurement it is strongly quantized as
multiple, evidently finite, steps
\cite{reilly}
while in the other it essentially vanishes
\cite{depic} as might be ideally expected.

\subsection{Negative Intrinsic Resistance}

A more recent set of four-point measurements by Gao {\em et al}.
\cite{gao}
shows that $R_4$ can even be negative. This might be
explained by the general B\"uttiker four-terminal resistance formula
\cite{butt}.
If so, however, this would also entail a set of contradictions
\cite{comment}:

\begin{itemize}
\item
if $R_4 < 0$ then ${\cal T} > 1$ and unitarity (probability)
is not conserved albeit the Landauer model is predicated
on unitary single-electron propagation.

\item
If $R_4 < 0$ then the power dissipation in the
device proper is $P = I^2 R_4 < 0$ and the device must spontaneously
be giving up energy to the rest of the circuit.
\end{itemize}

\noindent
The same dilemmas arise for yet another, three-terminal, experiment
\cite{kay}
announcing the measurement of ``absolute negative resistance''.
Their explanation is again in terms of the B\"uttiker
three-terminal formula (a restriction of the four-terminal relation).
Nevertheless all the transmission probabilities, being positive,
cannot produce a negative resistance from the three-terminal B\"uttiker
formula. Just as before the negative resistance in this experiment
directly violates the B\"uttiker theory.

The above are only a few examples of the counter-intuitive
consequences readily drawn from the accepted literature on
mesoscopic transport (taking on trust the integrity
of works that have been peer reviewed).
It is clear that various significant experiments on ballistic
conduction are in contradiction not only with prevailing theory
but also with one another, if not internally.

Thus it is not unfair
to ask how much of the content of certain published experiments
is actual fact and how much of it begs certain favoured questions,
having more the character of wish fulfilment than hard reality.
But such is not for us to analyze;
readers are free to form their own conclusions -- or not.

\section{Summary}

We have argued that a canonical kinetic approach to mesoscopic
transport provides -- uniquely -- a microscopic account
of conductance and noise in quantum point contacts.
It does so, and will {\em always} do so, free of gratuitous guesswork
and question-begging.

Unified models of quantum transport and fluctuations will yield,
and have yielded already, a natural and detailed understanding
of core non-equilibrium processes inside a quantum wire.
Fluctuations, by way of current noise,
carry much more information on the
internal dynamics of mesoscopic systems. Such knowledge
is not accessible through the current--voltage characteristics alone,
so that further device-noise experiments,
especially at constant current, would be of foremost importance.

To our mind, the heart and soul of a mesoscopic theory rest
with its physical integrity, and securing its credibility
requires sustained, diligent and harmonious work by
experimentalists and theorists.
But the theoretical task is not made any easier when
experimental works, appearing in the record, contradict
one another and even themselves -- not to mention
the entire spectrum of mesoscopic models (widely accepted or otherwise).

\end{document}